\documentclass[useAMS,usenatbib]{mn2e}
\usepackage{graphicx}
\usepackage{amssymb}
\usepackage{color}

\title[Non-variable TeV emission from the extended jet of a blazar]{Non-variable TeV emission from the extended jet of a blazar
in the stochastic acceleration scenario: the case of the hard TeV
emission of 1ES 1101-232}
\author[Yan et al.]{Dahai Yan, Houdun Zeng, and Li Zhang\thanks{E-mail: lizhang@ynu.edu.cn}\\
Department of Physics, Yunnan University, Kunming, China}

\begin{document}
\pagerange{\pageref{firstpage}--\pageref{lastpage}} \pubyear{2010}

\maketitle

\label{firstpage}

\begin{abstract}
The detections of X-ray emission from the kiloparsec-scale jets of
blazars and radio galaxies may imply the existence of high energy
electrons in these extended jets, and these electrons could
produce high energy emission through inverse Compton (IC) process.
In this paper we study the non-variable hard TeV emission from a
blazar. The multi-band emission consists of two components: one is
the traditional synchrotron self-Compton (SSC) emission from the
inner jet, and the other is the emission produced via SSC and IC
scattering of cosmic microwave background (CMB) photons (IC/CMB)
and extragalactic background light (EBL) photons (IC/EBL) by
relativistic electrons in the extended jet under the stochastic
acceleration scenario. Such a model is applied to 1ES 1101-232.
The results indicate that (1) the non-variable hard TeV emission
of 1ES 1101-232 can be reproduced well, which is dominated by
IC/CMB emission from the extended jet, using three characteristic
values of Doppler factor ($\delta_{\rm D}=5,10,15$) for the TeV
emitting region in the extended jet; and (2) in the cases of
$\delta_{\rm D}=15$ and 10, the physical parameters can achieve
the equipartition (or quasi-equipartition) between the relativistic electrons and the
magnetic field; In contrast, the physical parameters largely
deviate from the equipartition for the case of $\delta_{\rm D}=5$. It is therefore concluded that the TeV
emission region of 1ES 1101-232 in the extended jet should be
moderately or highly beamed.

\end{abstract}

\begin{keywords}
galaxies: BL Lacertae objects: individual: 1ES 1101-232 --
galaxies: active -- gamma-rays: general -- radiation mechanisms:
non-thermal
\end{keywords}

\section{Introduction}

Blazars are the most extreme class of active galactic nuclei
(AGNs). Their spectral energy distributions (SEDs) are
characterized by two distinct bumps: the first bump located at
low-energy band is dominated by the synchrotron emission of
relativistic electrons in a relativistic jet, and the second bump
located at high-energy band could be produced by IC scattering
(e.g., B\"{o}ttcher 2007). Various soft photon sources seed SSC
process (e.g., Rees 1967; Maraschi et al. 1992) and external
Compton (EC) process (e.g., Dermer \& Schlickeiser 1993; Sikora et
al. 1994) in the jet to produce $\gamma$-rays. Hadronic models
have also been proposed to explain the multi-band emissions of
blazars (e.g., Mannheim 1993; M\"{u}cke et al. 2003).

More than 40 blazars have been detected in TeV band and most of
them are high frequency-peaked BL Lac objects (HBLs). GeV - TeV
photons from blazars would be absorbed by interaction with EBL
photons, through the pair-production process (e.g.,
\cite{Stecker92}). Therefore, the observed high energy emission
from a TeV blazar must be less luminous and its spectrum must be
steeper compared with its intrinsic emission. A large number of
observational evidences support that the TeV emissions of HBLs can
be explained by the SSC radiation inside their jets. However, the
recent observations of TeV emission from some HBLs challenge the
classic SSC scenario. Indeed, the intrinsic TeV spectra of some
relatively distant sources (e.g. 1ES 0229+200, z = 0.14,
\cite{Aharonian07a}; 1ES 1101-232, z = 0.186, \cite{Aharonian06};
1ES 0347-121, z=0.188, \cite{Aharonian07b}) are very hard (the
intrinsic TeV spectral index $\Gamma_{\rm int}\sim 1.5$), even
considering the low level EBL models (e.g.,
\cite{Franceschini,finke10}). The standard shock acceleration
theories predict the particle distribution index $p\geq2$, where
$n(\gamma)=\gamma^{-p}$. This would correspond to a limiting
intrinsic photon spectral index $\Gamma_{\rm int}\geq1.5$.
Moreover, as the suppression of the cross-section due to
Klein-Nishina (KN) effect becomes important in the TeV emission
from HBLs, steeper intrinsic photon spectra would be expected.

Many possible solutions have been proposed to overcome this
problem of hard TeV emission. \citet{kata06a} suggested that the
hard TeV emission of 1ES 1101-232 could be explained in SSC
scenario by assuming a narrow high energy electrons distribution
with a large value of minimum energy cutoff ($\sim10^5$). As shown
in \citet{tavecc09}, this solution might be supported by the
simultaneous UV and X-ray observations of 1ES 0229+200 performed
by the {\it Swift} satellite. While this scenario would require
that emission process should be very inefficient but acceleration
process should be very efficient. The formation of such hard TeV
emission also would be achieved in an internal absorption scenario
\citep{Aharonian08,Zach}. Very recently, \citet{lefa11a} suggested
that considering either adiabatic losses or relativistic
Maxwell-like distributions formed by a stochastic acceleration is
an alternative option to create hard TeV spectra. \citet{lefa11b}
shown that the relative hard $\gamma$-ray spectrum of Mkn 501
could be formed in a multi-zone model in the stochastic
acceleration scenario. The secondary gamma-rays produced
relatively close to the Earth by the interactions of cosmic ray
protons with background photon can also explain the hard TeV
emission of blazars (e.g., \cite{Essey}).

Previous works usually considered the emissions from the inner
jets (sub-parsec scale) of blazars. Alternatively, \citet{bott08}
suggested that the slowly variable, hard TeV emission of 1ES
1101-232 might be created via Compton up-scattering of CMB photons
by shock-accelerated electrons in an extended jet (kiloparsec
scale). X-ray emissions from the extended jets of blazars have
been detected (e.g., \cite{tavecc07,Sambruna,Massaro}). These evidences imply the
presence of high energy electrons in their extended jets. In such
an extended jet, the CMB photons would be dominant soft photons
for IC process (e.g., \cite{tavecc00}), and the emission of
Compton up-scattering of the EBL photons off relativistic
electrons may be also important at TeV band
\citep{Georganopoulos}, due to the Doppler effect. Moreover,
recent studies indicated that electrons in the large-scale jets of
AGNs may be accelerated to high energies by stochastic
acceleration (e.g., \cite{fan,O'Sullivan}). The possible TeV
emissions from the extended jets of some radio galaxies have been
explored (e.g., \cite{Hardcastle11}). In the AGN unified scheme
\citep{Urry}, it is natural and worthwhile to wonder whether the
kind of non-variable TeV emissions of some blazars can be produced
at least in part in their extended jets.

In this paper, we study possible TeV emission from the extended
jet of a blazar. We assume that multi-band emission from a
non-variable TeV blazar includes two components: one is from the
inner jet and the other is from the extended jet. In this model,
the emission from the inner jet is produced in a conventional SSC
model; and the emission from the extended jet is produced in a
self-consistent SSC+IC/CMB/EBL model under the stochastic
acceleration scenario. We apply the model to explain the
non-variable hard TeV emission of 1ES 1101-232. The cosmological
parameters ($H_0, \Omega_m, \Omega_{\Lambda}$) = (70 km s$^{-1}$
Mpc$^{-1}$, 0.3, 0.7) are used throughout this paper.

\section{The model}

As mentioned above, we assume that the emission from a
non-variable TeV blazar is produced at both inner and extended
jet. In this model, non-thermal photons are produced by both the
synchrotron radiation and IC scattering of relativistic electrons
in a spherical blob which is moving relativistically at a small
angle to our line of sight, and the observed radiation is strongly
boosted by a relativistic Doppler factor $\delta_{\rm D}$. We give
the specific descriptions of this model below.

\subsection{The conventional SSC model in the inner jet}

We use the conventional SSC model to produce the emission from the
inner jet. Here we assume a broken power-law electrons energy
distribution and use the relativistic electron distribution given
by \citet{dermer09}, i.e.,
\begin{eqnarray}
N_{\rm e}^{\prime}(\gamma^{\prime})=K_{\rm
e}^{\prime}H(\gamma^{\prime};\gamma_{\rm min}^{\prime},\gamma_{\rm
max}^{\prime})\{{\gamma^{\prime
-p_1}\exp(-\gamma^{\prime}/\gamma_{\rm b}^{\prime})}
\nonumber \\
\times H[(p_{\rm 2}-p_{\rm 1})\gamma_{\rm
b}^{\prime}-\gamma^{\prime}]+[(p_{\rm 2}-p_{\rm 1})\gamma_{\rm
b}^{\prime}]^{p_{\rm 2}-p_{\rm 1}}\gamma^{\prime -p_{\rm 2}}
\nonumber \\
\times \exp(p_{\rm 1}-p_{\rm 2})H[\gamma^{\prime}-(p_{\rm
2}-p_{\rm 1})\gamma_{\rm b}^{\prime}]\},
\end{eqnarray}
where $K_{\rm e}^{\prime}$ is the normalization factor, in unit of
$\rm cm^{-3}$. $H(x;x_{1},x_{2})$ is the Heaviside function:
$H(x;x_{1},x_{2})=1$ for $x_{1}\leq x\leq x_{2}$ and
$H(x;x_{1},x_{2})=0$ everywhere else; as well as $H(x)=0$ for
$x<0$ and $H(x)=1$ for $x\geq0$. The minimum and maximum energies
of electrons in the blob are $\gamma_{\rm min}^{\prime}$ and
$\gamma_{\rm max}^{\prime}$, respectively. The spectrum is
smoothly connected with indices $ p_1$ and $p_2$ below and above
the electron's break energy $\gamma_{\rm b}^{\prime}$. Note that
here and throughout the paper, unprimed quantities refer to the
distant observer's frame on the Earth
and primed ones refer to the co-moving frame.

Then, the synchrotron flux is calculated as (Finke et al. 2008)
\begin{equation}
\nu F^{\rm syn}_{\nu}=\frac{\sqrt{3}\delta^4_{\rm
D}\epsilon'e^3B^{\prime}}{4\pi h d^2_{\rm
L}}\int^\infty_0d\gamma'N'_e(\gamma')(4\pi R_{\rm b}^{\prime
3}/3)R(x)\;\;, \label{syn}
\end{equation}
where $e$ is the electron charge, $B^{\prime}$ is the magnetic
field strength, $R_{\rm b}^{\prime}$ is blob's radius, $h$ is the
Planck constant, and $d_{\rm L}$ is the distance to the source
with a redshift $z$. Here $m_{\rm
e}c^2\epsilon^{\prime}=h\nu(1+z)/\delta_{\rm D}$ is synchrotron
photons energy in the co-moving frame, where $m_{\rm e}$ is the
rest mass of electron and $c$ is the speed of light. Here we use
an approximation for $R(x)$ given by Finke et al. (2008). The
synchrotron spectral energy density is
\begin{equation}
u^{\prime}_{\rm syn}(\epsilon^{\prime})=\frac{R_{\rm
b}^{\prime}}{c}\frac{\sqrt{3}e^3B^{\prime}}{h}\int^\infty_0d\gamma'N'_e(\gamma')R(x)\;.
\label{usyn}
\end{equation}

The SSC flux $\nu F_{\nu}$ is given by Finke et al. (2008)
\begin{eqnarray}
\nu F^{\rm SSC}_{\nu}=\frac{3}{4}c\sigma_{\rm
T}\epsilon_{s}^{\prime 2}\frac{\delta_{\rm D}^4}{4\pi d_{\rm
L}^2}\int_0^{\infty}
d\epsilon^{\prime}\frac{u^{\prime}_{\rm syn}(\epsilon^{\prime})}{\epsilon^{\prime 2}} \nonumber \\
\times\int_{\gamma_{\rm min}^{\prime}}^{\gamma_{\rm max}^{\prime}}
d\gamma^{\prime}\; \frac{N^{\prime}_e(\gamma^{\prime})(4\pi R_{\rm
b}^{\prime 3}/3)}{\gamma^{\prime 2}}F_{\rm
C}(q^{\prime},\Gamma_{\rm e}^{\prime})\;, \label{ssc}
\end{eqnarray}
where $\sigma_{\rm T}$ is the Thomson cross section, $m_{\rm
e}c^2\epsilon^{\prime}_{s}=h\nu(1+z)/\delta_{\rm D}$ is the energy
of IC scattered photons in the co-moving frame, $F_{\rm
C}(q^{\prime},\Gamma_{\rm e}^{\prime})=2q^{\prime}{\rm
ln}q^{\prime}+(1+2q^{\prime})(1-q^{\prime})+\frac{q^{\prime
2}\Gamma_{\rm e}^{\prime 2}}{2(1+q^{\prime}\Gamma_{\rm
e}^{\prime})}(1-q^{\prime})$, $
q^{\prime}=\frac{\epsilon^{\prime}/\gamma^{\prime}}{\Gamma^{\prime}_{\rm
e}(1-\epsilon^{\prime}/\gamma^{\prime})}$, $\Gamma_{\rm
e}^{\prime}=4\epsilon^{\prime}\gamma^{\prime}$, and $
\frac{1}{4\gamma^{\prime 2}}\leq q^{\prime}\leq1$.

\subsection{The self-consistent SSC+IC/CMB/EBL model in the extended jet under the
stochastic acceleration scenario}

The self-consistent SSC+IC/CMB/EBL model including a acceleration process in the stochastic
acceleration scenario is used to create the emission from the
extended jet. A physical self-consistent description of stochastic
acceleration in a time evolution scenario can be achieved through
a kinetic equation, and the kinetic equation is given as
(\cite{kata06b}; also see \cite{Tramacere11} and
\cite{Weidinger10})
\begin{eqnarray}
\frac{\partial N^{\prime}(\gamma^{\prime},t)} {\partial t}  =
\frac{\partial}{\partial \gamma^{\prime}}
\left[\left\{C^{\prime}(\gamma^{\prime},t)-A^{\prime}(\gamma^{\prime},t)\right\}N^{\prime}(\gamma^{\prime},t)\right] \nonumber \\
+ \frac{\partial}{\partial
\gamma^{\prime}}\left[D^{\prime}(\gamma^{\prime},t) \frac{\partial
N^{\prime}(\gamma^{\prime},t)}{\partial \gamma^{\prime}} \right] -
E^{\prime}(\gamma^{\prime},t) + Q^{\prime}(\gamma^{\prime},t)\ ,
\label{eed}
\end{eqnarray}
where $D^{\prime}(\gamma^{\prime},t)$ is the momentum diffusion
coefficient and
$A^{\prime}(\gamma^{\prime},t)=(2/\gamma^{\prime})D^{\prime}(\gamma^{\prime},t)$
is the average energy change term resulting from the
momentum-diffusion process. In this work we assume that the
acceleration timescale $t_{\rm acc}$ and escape timescale $t_{\rm
esc}$ are both independent of electron energy, which is the case
of the hard sphere approximation. Hence,
$D^{\prime}(\gamma^{\prime})=(1/2t_{\rm acc})\gamma^{\prime 2}$,
$A^{\prime}(\gamma^{\prime})=\gamma^{\prime}/t_{\rm acc}$ and the
escape term $E^{\prime}(\gamma^{\prime})=
N^{\prime}(\gamma^{\prime},t)/t_{\rm esc}$.
$Q^{\prime}(\gamma^{\prime},t)$ is the injection term.
$C^{\prime}(\gamma^{\prime},t)$ is the total cooling rate. In
addition to the synchrotron cooing rate
$(\frac{d\gamma^{\prime}}{dt})_{\rm syn}$ and SSC cooling rate
$(\frac{d\gamma^{\prime}}{dt})_{\rm SSC}$, we consider the IC/CMB
cooling rate $(\frac{d\gamma^{\prime}}{dt})_{\rm IC/CMB}$ and
IC/EBL cooling rate $(\frac{d\gamma^{\prime}}{dt})_{\rm IC/EBL}$.
Therefore,
$C^{\prime}(\gamma^{\prime},t)=(\frac{d\gamma^{\prime}}{dt})_{\rm
syn}+(\frac{d\gamma^{\prime}}{dt})_{\rm
SSC}+(\frac{d\gamma^{\prime}}{dt})_{\rm
IC/CMB}+(\frac{d\gamma^{\prime}}{dt})_{\rm IC/EBL}$. For
synchrotron cooling, $(\frac{d\gamma^{\prime}}{dt})_{\rm
syn}=\frac{4\sigma_{\rm T}c}{3m_{\rm e}c^2}U^{\prime}_{\rm
B}\gamma^2$, where $U^{\prime}_{\rm B}=\frac{B^{\prime 2}}{8\pi}$
is the magnetic field energy density. The SSC cooling rate using
the full KN cross section (e.g., \cite{Jones68,bott97,finke08}) is
\begin{equation}
(\frac{d\gamma^{\prime}}{dt})_{\rm SSC} = \frac{3 \sigma_{\rm
T}}{8m_{\rm e}c}\int^{\infty}_{0}d\epsilon^{\prime}
             \frac{u_{\rm syn}^\prime(\epsilon^{\prime} )}{\epsilon^{\prime 2}}\ G(\gamma^{\prime}\epsilon^{\prime})\,
\end{equation}
where $G(E) =
\frac{8}{3}E\frac{1+5E}{(1+4E)^2}-\frac{4E}{1+4E}\left(\frac{2}{3}+\frac{1}{2E}+\frac{1}{8E^2}\right)$
$+\ln(1+4E)\left(
1+\frac{3}{E}+\frac{3}{4}\frac{1}{E^2}+\frac{\ln[1+4E]}{2E}-\frac{\ln[4E]}{E}
\right)$ $-\frac{5}{2}\frac{1}{E}\ +\
\frac{1}{E}\sum^{\infty}_{n=1}
\frac{(1+4E)^{-n}}{n^2}-\frac{\pi^2}{6E}-2,$ and $u_{\rm
syn}^\prime (\epsilon^{\prime}) $ is given by equation
(\ref{usyn}). We calculate the IC/CMB and IC/EBL cooling rates
using the method given by \citet{Moderski05}, which fully takes
into account KN effects,
\begin{equation}
(\frac{d\gamma^{\prime}}{dt})_{\rm IC/CMB/EBL}=\frac{4\sigma_{\rm
T}c}{3m_{\rm e}c^2}\gamma^{\prime 2}F_{\rm KN}\ ,
\end{equation}
where $F_{\rm KN}=\int^{\epsilon^{\prime}_{\rm
max}}_{\epsilon^{\prime}_{\rm min}}f_{\rm
KN}(4\gamma^{\prime}\epsilon^{\prime})u^{\prime}_{\rm
CMB/EBL}(\epsilon^{\prime})d\epsilon^{\prime}$. $f_{\rm
KN}(x)\simeq\frac{9}{2x^2}({\rm ln}x-11/6)$ for $x\gg1$. $f_{\rm
KN}(x)$ can be approximated by $f_{\rm KN}(x)\simeq1/(1+x)^{1.5}$
for $x\lesssim10^4$. For the case of the CMB radiation in the
stationary frame of the host galaxy, $u_{\rm
CMB}(\epsilon)=\frac{8\pi m_{\rm e}c^2(m_{\rm
e}^3c^3)}{h^3}\frac{\epsilon^3}{{\rm exp}(\epsilon/\Theta)-1}$,
where $\Theta=\kappa_{\rm B}T/m_{\rm e}c^2$ is the dimensionless
temperature of the blackbody radiation field, $T=2.72(1+z)$ K and
$\kappa_{\rm B}$ is the Boltzmann constant. We use the spectral
EBL energy density expected in \citet{finke10}. The spectral
energy densities in the co-moving frame transform as
$u^{\prime}(\epsilon^{\prime})=\delta_{\rm
D}u(\epsilon^{\prime}/\delta_{\rm D})$ (e.g.,\citet{G09}).

\begin{figure}
\begin{center}
\includegraphics[width=90mm]{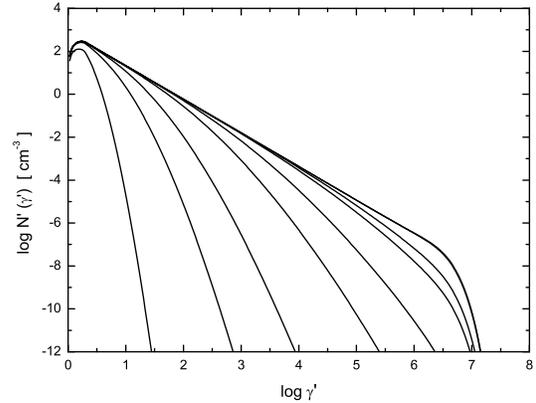}
\end{center}
\caption{The reproduced electron spectra in Fig.~4 of
\citet{kata06b} by using our code.}
   \label{Fig1}
\end{figure}

To obtain the self-consistent relativistic electron energy
distribution, we must solve equation (\ref{eed}) numerically. In
the numerical calculations, we adopt the numerical method given by
\citet{Park}, which is firstly proposed by \citet{cc}. The method
is a finite difference scheme, which uses the centered difference
of the diffusive term, and a weighted difference for the advective
term. We have carefully tested our code by running it with
parameters identical to those used for Fig.~4 of \citet{kata06b},
only considering the synchrotron and SSC radiative cooling. We
find good agreement with their results. The reproduced electron
spectra in Fig.~4 of \citet{kata06b} using our code are shown in
our Fig.~\ref{Fig1}.

After obtaining the relativistic electron energy distribution, we
can calculate the synchrotron-SSC flux and IC/CMB/EBL flux from
the extended jet by using equation (\ref{syn}) and equation
(\ref{ssc}) with the corresponding electron spectrum and energy
densities of soft photons field.

High energy emission from both the inner and extended jet would be
modified by a factor of $\exp(-\tau)$ due to the absorption of
EBL. In this paper, we use the optical depth $\tau$ expected in
\citet{finke10}. This EBL model is a low level one, which is
consistent with the widely used one (e.g., \cite{Franceschini})
and the newly proposed one (e.g., \cite{Dom2011}).

\begin{table*}
\caption{The modeling parameters for the inner and extended jets.
} \label{para}
\begin{center}
\begin{tabular}{lccccccccc}
 \hline\hline
inner jet & $B^{\prime}$ & $\gamma^{\prime}_{\rm min}$ &
$\gamma^{\prime}_{\rm max}$ & $\gamma^{\prime}_{\rm b}$ &
$\delta_{\rm D}$ & $K^{\prime}_{\rm e}$ & $p_1$ & $p_2$ &
$R^{\prime}_{\rm b}$\\
          & (G) & & & & & $({\rm cm}^{-3})$ & & & (cm) \\\hline
& 0.1 & 1.0 & $6.0\times10^5$ & $2.0\times10^5$ & 24.7 & 19.8 & 2.0 &
4.0 & $7.5\times10^{16}$
\\\hline\hline
extended jet & $B^{\prime}$ ($\mu$G) & $Q^{\prime}\ ({\rm cm^{-3}\ s^{-1}})$ & $R^{\prime}_{\rm b}$ (cm) & $t_{\rm esc}$ (s) & $t_{\rm acc}/t_{\rm esc}$ & $\delta_{\rm D}$ & $U^{\prime}_{\rm e}/U^{\prime}_{\rm B}$\\
\hline
highly beamed & 23.0 & $8.0\times10^{-20}$ & $5.0\times10^{20}$ & $R^{\prime}_{\rm b}/c$ & 0.86 & 15 & 1.0\\
moderately beamed & 58.5 & $5.0\times10^{-19}$ & $5.0\times10^{20}$ & $R^{\prime}_{\rm b}/c$ & 0.86 & 10 & 1.0\\
mildly beamed & 29.3 & $1.0\times10^{-17}$ & $5.0\times10^{20}$ & $R^{\prime}_{\rm b}/c$ & 0.78 & 5 & 112.3\\
  \hline
\end{tabular}
\end{center}
\end{table*}

\begin{table*}
\caption{The radiative ($P_{\rm r}$), Poynting flux
($P_{\rm B}$), kinetic ($P_{\rm e}$  and
$P_{\rm p}$), and injection ($L_{\rm inj}$) powers for the inner and extended jets. } \label{power}
\begin{center}
\begin{tabular}{lccccccccc}
 \hline\hline
inner jet & $P_{\rm r}$ ($\rm erg\ s^{-1}$) & $P_{\rm B}$ ($\rm erg\ s^{-1}$) & $P_{\rm e}$ ($\rm erg\ s^{-1}$) & $P_{\rm p}$ ($\rm erg\ s^{-1})$\\
\hline
& $4.1\times10^{43}$ & $1.3\times10^{44}$ & $5.4\times10^{43}$ & $9.2\times10^{45}$\\

\hline\hline
extended jet & $P_{\rm r}$ ($\rm erg\ s^{-1}$) & $P_{\rm B}$ ($\rm erg\ s^{-1}$) & $P_{\rm e}$ ($\rm erg\ s^{-1}$) & $P_{\rm p}$ ($\rm erg\ s^{-1}$) & $L_{\rm inj}$ ($\rm erg\ s^{-1}$)\\
\hline
highly beamed & $2.5\times10^{43}$ & $1.1\times10^{44}$ & $1.1\times10^{44}$ & $1.5\times10^{43}$ & $5.2\times10^{40}$\\
moderately beamed & $1.2\times10^{44}$ & $3.2\times10^{44}$ & $3.2\times10^{44}$ & $4.1\times10^{43}$ & $1.4\times10^{41}$\\
mildly beamed & $2.4\times10^{44}$ & $2.0\times10^{43}$ & $2.3\times10^{45}$ & $2.0\times10^{44}$ & $7.2\times10^{41}$\\
  \hline
\end{tabular}
\end{center}
\end{table*}

\section{Application to 1ES 1101-232}

1ES 1101-232 resides in an elliptical host galaxy (e.g.,
\cite{Remillard}). An extended jet structure at a few kiloparsec
distance from the core is revealed by the radio observations
(\cite{LM}). The TeV observations of 1ES 1101-232 were performed
by the H.E.S.S. Cherenkov telescopes in April and June 2004, and
in March 2005. Simultaneous observations were carried out with
H.E.S.S., X-ray measurements with RXTE, and optical measurements
with the ROTSE 3c robotic telescope in March 2005, following the
detection of a weak signal in the 2004 H.E.S.S. observations
\citep{Aharonian07c}. No TeV variability was found in these
observations, and a moderate flux changes were observed with RXTE
in March 2005 \citep{Aharonian07c}. No very significant
signal from 1ES 1101-232 is detected by {\it Fermi}-LAT in its
first year scientific operation \citep{abdoagn}, and only the
flux upper limits derived from the first year {\it Fermi}-LAT
observations are available (e.g., \cite{Neronov,Costamante2011}).
The two years scientific operation of {\it Fermi}-LAT
reports a significant detection of 1ES 1101-232 at a significance
5.2$\sigma$ with the photon spectral index $\Gamma=1.80\pm0.31$
\citep{Ackermann}. More significant signals ($\sigma \sim10$) from
1ES 1101-232 have been detected in {\it Fermi}-LAT's 3.5 years
observations \citep{finke12}. We will use {\it Fermi}-LAT 3.5
years average spectrum. Since the detections made by H.E.S.S. in
2004 are not significant, we select the simultaneous multi-band
data derived in the observations in March 2005 in this paper.

The variability at optical and X-ray frequencies of 1ES 1101-232
have been detected (e.g., \cite{Remillard,Romero,Wolter}).
Moreover, the IC/CMB process having a radiative cooling timescale
$\sim10^3$ years (\cite{bott08}), such long cooling timescale
indicates that IC/CMB emission, as well as the synchrotron
emission associated with the same high energy electrons and
emission region, will be slowly variable. In addition, the very
large size of the emission region in the extended jet (e.g.,
\cite{tavecc07,fan}) also indicates that emission from such a
region should be slowly variable. We therefore argue that the
moderately variable optical-X-ray radiation of 1ES 1101-232 should be dominated by the emissions originating in a compact region in the inner jet.

As discussed above, we use the conventional SSC model in the inner
jet to produce the optical-X-ray emission. The self-consistent
SSC+IC/CMB/EBL model is used to create the TeV emission from the
extended jet. In the extended jet, we assume that electrons with
energy $\gamma^{\prime}_{\rm inj}\approx27$ are continuously
injected into a blob with a constant injection rate $Q^{\prime}$.
The minimum and maximum energies of electrons in this blob are set
as $\gamma^{\prime}_{\rm min}=1.0$ and $\gamma^{\prime}_{\rm
max}=10^7$.

The Doppler effect is crucial
for the IC/CMB/EBL emission because it can affect the CMB energy
density and the CMB photons energy in the co-moving frame, and
inversely it affects the relativistic electrons energy density
required to produce the TeV emission. Unfortunately, Doppler
factor is poorly constrained. It is possible that the jets of
blazars remain highly relativistic at kiloparsec scales (e.g.,
\cite{Atoyan,Jorstad}). However, recent very long baseline
interferometry (VLBI) observations indicate that the bulk Lorentz
factors in the parsec-scale radio jets of TeV HBLs are fairly low,
$\Gamma<5$ (e.g.,\cite{Piner08,Piner10}). Because of the
importance of Doppler effect, we assume three values of Doppler
factor for the TeV emission region in the extended jet, which are
$\delta_{\rm D}=15,10,5$ corresponding to the highly beamed,
moderately beamed and mildly beamed cases respectively, to
reproduce the TeV emission of 1ES 1101-232.
In the following, we show the results of reproducing
SED.

\begin{figure}
\begin{center}
\includegraphics[width=90mm]{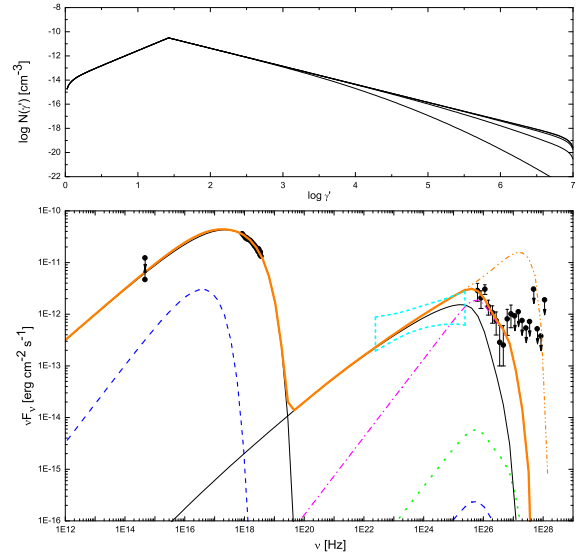}
\end{center}
\caption{The case of highly beamed extended jet with $\delta_{\rm
D}=15$. Upper panel: the evolution of the electron spectrum in the
extended jet. From left to right: t= ( 3, 5, 7, 9, 13) $t_{\rm
acc}$. Lower panel: SED of 1ES 1101-232. The filled circles are
the simultaneous optical - X-ray - TeV data taken from
\citet{Aharonian07c}. The area surrounded by short-dashed line is
the {\it Fermi}-LAT 3.5 years average spectrum integrated over
August 2008 to February 2012 \citep{finke12}. The curves represent
synchrotron-SSC emission from the inner jet (thin solid line),
synchrotron-SSC emission from the extended jet (dashed line),
IC/EBL emission from the extended jet (dotted line) and IC/CMB
emission from the extended jet (dot-dashed line). The thick solid
line is the total emission from the inner and extended jet with
absorption of EBL and the double-dot-dashed line is the intrinsic
total spectrum without absorption of EBL.}
   \label{case15}
\end{figure}

The case of highly beamed extended jet with $\delta_{\rm D}=15$:
the results are shown in Fig.~\ref{case15}. It can be seen that
when $t>9\ t_{\rm acc}$, the electron spectrum produced by the
stochastic acceleration tends to be steady. This distribution in
steady-state develops a rising low-energy power-law tail (left
side of $\gamma^{\prime}_{\rm inj}$ in upper panel of
Fig.~\ref{case15}) with the theoretic spectral index
$n\simeq-2-\frac{t_{\rm acc}}{2t_{\rm esc}}$ (the acceleration
theory predicts $n\simeq-2.43$, and our numerical calculations
give $n\approx-2.49$), but at high energies it presents a
power-law spectrum with an exponential cut-off (right side of
$\gamma^{\prime}_{\rm inj}$ in upper panel of Fig.~\ref{case15}).
The spectral index of power-law distribution formed in the
stochastic acceleration scenario at
$\gamma^{\prime}>\gamma^{\prime}_{\rm inj}$ is
$n\simeq1+\frac{t_{\rm acc}}{2t_{\rm esc}}$ (the acceleration
theory predicts $n\simeq1.43$, and our numerical calculations give
$n\approx1.49$), which differs from the spectral index
$n\simeq1+\frac{t_{\rm acc}}{t_{\rm esc}}$ that obtained from the
basic shock acceleration (e.g., \cite{Kirk,kata06b}). The cut-off
energy can be obtained when the cooling time scale $t_{\rm cool}$
satisfies the condition $t_{\rm cool} (\gamma^{\prime}_{\rm
c})=t_{\rm acc}$. We adopt this steady-state electron spectrum to
produce the TeV emission from the extended jet. As shown in lower
panel of Fig.~\ref{case15}, a good representation of SED can be
achieved. The synchrotron emission from the inner jet contributes
to the optical-X-ray emission of 1ES 1101-232. At GeV band, SSC
emission from the inner jet is dominant. The parameters we used
are appropriate for the emission region in the inner jet with the
ratio between the relativistic electrons energy density and the
magnetic field energy density, $U^{\prime}_{\rm e}/U^{\prime}_{\rm
B}\approx0.4$. The parameters are listed in Table \ref{para}. The
IC/CMB emission from the extended jet dominates the TeV emission,
while IC/EBL emission as well as the SSC emission from the
extended jet are negligible in this case.  Our results indicate
that the intrinsic TeV flux peak is located at $\sim$ 4\ TeV, and
the intrinsic TeV spectral index is $\Gamma_{\rm int}\sim1.5$,
which are consistent with that inferred by \citet{Aharonian07c}.
As required, the injected power ($L_{\rm inj}=\frac{4}{3}\pi
R^{\prime 3}_{\rm b}\frac{\delta^2_{\rm
D}}{4}Q^{\prime}\gamma^{\prime}_{\rm
inj}\approx5.2\times10^{40}\rm\ erg\ s^{-1}$ \citep{Weidinger11})
in the stationary frame of the host galaxy for the extended jet
(see Table \ref{power}) exceeds the observed radio power of the
extended radio structure at 1.5 GHz, $L^{\rm
ext}_{1.5}\approx3.8\times10^{40}\ {\rm erg\ s^{-1}}$ \citep{LM}.
Because of the highly relativistic extended jet, the required
energy density of relativistic electrons is small,
$U^{\prime}_{\rm e}\approx2.1\times10^{-11}\ {\rm erg\ cm^{-3}}$.
The parameters we adopted for the extended jet can achieve the
equipartition condition between the relativistic electrons and the
magnetic field ($U^{\prime}_{\rm e}/U^{\prime}_{\rm
B}\approx1.0$), and the synchrotron emission from the extended jet
is about one order of magnitude below the emission from the inner
jet.

\begin{figure}
\begin{center}
\includegraphics[width=90mm]{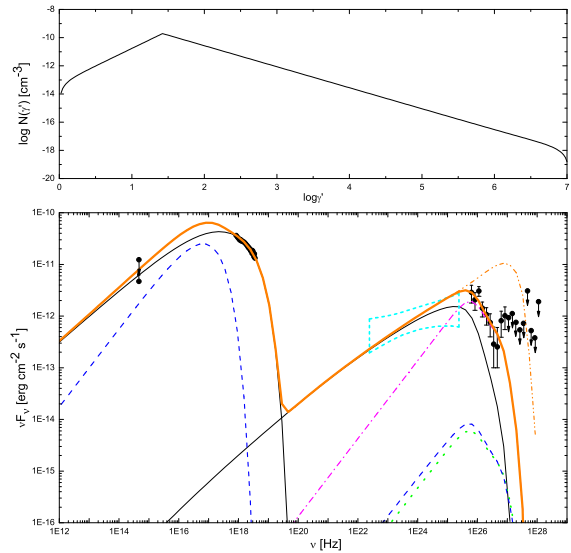}
\end{center}
\caption{The case of moderately beamed extended jet with
$\delta_{\rm D}=10$. Upper panel: the electron spectrum in steady
state. Lower panel: SED of 1ES 1101-232. The symbols and curves
are the same as that in Fig.~\ref{case15}.}
   \label{case10}
\end{figure}

The case of moderately beamed extended jet with $\delta_{\rm
D}=10$: the results are presented in Fig.~\ref{case10}. In this
case, we only present the electron spectrum in the steady state,
which is used to reproduce the TeV emission from the extended jet,
since the evolution of this electron spectrum is the same as that
in the former case. The emissions from the inner jet are the same
as that in the former case, and TeV emission is also dominated by
IC/CMB emission. In this case, the required energy density of
relativistic electrons becomes larger, $U^{\prime}_{\rm
e}\approx1.4\times10^{-10}\ {\rm erg\ cm^{-3}}$. In order to
achieve the equipartition with $U^{\prime}_{\rm e}/U^{\prime}_{\rm
B}\approx1.0$, a larger magnetic field strength ($U^{\prime}_{\rm
B}\sim 2U^{\prime}_{\rm CMB}$) is needed, which causes the
significant synchrotron emission from the extended jet. If
a magnetic field in quasi-equipartition with the relativistic
electrons with $U^{\prime}_{\rm e}/U^{\prime}_{\rm B}\sim 2.0$ is
used, the contribution of synchrotron emission from the extended
jet will become slight to the optical-X-ray emission of 1ES
1101-232.

\begin{figure}
\begin{center}
\includegraphics[width=90mm]{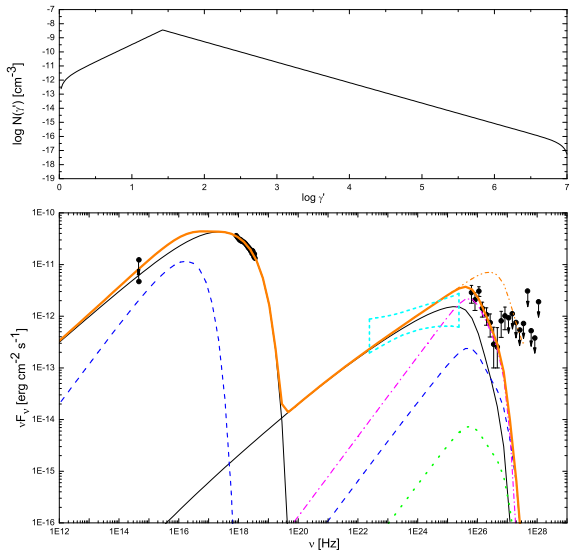}
\end{center}
\caption{The case of mildly beamed extended jet with $\delta_{\rm
D}=5$. Upper panel: the electron spectrum in steady state. Lower
panel: SED of 1ES 1101-232. The symbols and curves are the same as
that in Fig.~\ref{case15}.}
   \label{case5}
\end{figure}

The case of mildly beamed extended jet with $\delta_{\rm D}=5$: we
show the results in Fig.~\ref{case5}. The emissions from the inner
jet are still unchanged. In this case, for reproducing the TeV
emission well, a high energy density of relativistic
electrons in the extended jet is needed, $U^{\prime}_{\rm
e}\approx3.8\times10^{-9}\ {\rm erg\ cm^{-3}}$.  Even a large
magnetic field strength is used ($U^{\prime}_{\rm B}\sim
2U^{\prime}_{\rm CMB}$), the parameters with $U^{\prime}_{\rm e}/U^{\prime}_{\rm
B}\approx112.3$ still largely deviate from the equipartition.

We now can estimate the powers in the forms of radiation
($P_{\rm r}$), Poynting flux ($P_{\rm B}$), relativistic electrons
($P_{\rm e}$) and protons ($P_{\rm p}$) for the inner and extended
jet, respectively, which are all in the stationary frame of the
host galaxy.  All the powers are calculated as
\citep{Celotti,Celotti08}
\begin{equation}
P_{\rm i}=\pi R^{\prime 2}_{\rm b}\Gamma^2U^{\prime}_{\rm i}c\ ,
\end{equation}
where $U^{\prime}_{\rm i}$ (i=r, e, B, p) are the energy densities
associated with the radiation $U^{\prime}_{\rm r}$, the emitting
electrons $U^{\prime}_{\rm e}$, magnetic field $U^{\prime}_{\rm
B}$ and protons $U^{\prime}_{\rm p}$ in the comoving frame. We
calculate $U^{\prime}_{\rm p}$ by assuming one proton per emitting
electron, then $U^{\prime}_{\rm p}=U^{\prime}_{\rm e}(m_{\rm
p}/m_{\rm e})/\langle\gamma^{\prime}\rangle$ \citep{Celotti08},
where $\langle\gamma^{\prime}\rangle$ is the average energy of
relativistic electrons. Here, we take the bulk Lorentz factor
$\Gamma=\delta_{\rm D}$. The power carried in the form of the
produced radiation can be rewritten as by using  $U^{\prime}_{\rm
r}=L^{\prime}/(\pi R^{\prime 2}_{\rm b}c)$ \citep{Celotti08}
\begin{equation}
P_{\rm}=L^{\prime}\Gamma^2\approx L\frac{\Gamma^2}{\delta^4_{\rm D}}\ ,
\end{equation}
where $L$ is total no-thermal luminosity.

Our predicted powers are listed in Table \ref{power}. For
the inner jet, we have that $P_{\rm r}\sim P_{\rm e}\le P_{\rm
B}$. This follows the fact that the synchrotron luminosity from
the inner jet is larger than the $\gamma$-ray one. For the
extended jets with $\delta_{\rm D}=10, 15$, we have that $P_{\rm
r}\sim P_{\rm e}\sim P_{\rm B}$. Since $P_{\rm r}\sim P_{\rm e}$,
relativistic electrons cannot be the primary energy carriers in
the jet \citep{Celotti08}. Under the assumption of one proton per
emitting electron, our results indicate that a substantial
fraction of the total jet power in the inner jet is carried by
protons. However, it can be seen that in the extended jet, the jet
power in protons is far smaller than the one in emitting
electrons, with $P_{\rm p}/P_{\rm e}\sim0.1$, which contradicts
the scenario that a sizeable fraction of jet power is carried out
to large distances by the protons content of the jet. Moreover,
Poynting flux in the inner jet is comparable with the ones in the
extended jets for the cases of $\delta_{\rm D}=10, 15$. Therefore,
relativistic leptons as well as Poynting flux also cannot be the
primary energy carriers. It is likely that for the case of 1ES
1101-232 the cold electron positron pairs are the primary energy
carriers in the jet (e.g., \citet{Celotti08}).

As mentioned in introduction, \citet{bott08} have modeled
the TeV emission from 1ES 1101-232 in the frame of Compton
up-scattering of CMB photons by shock-accelerated electrons in an
extended jet with $\delta_{\rm D}=15$. It is worthwhile to compare their
results with ours. For the extended jet with $\delta_{\rm D}=15$,
the injected power in our model is $L_{\rm inj}= 5.2\times
10^{40}\rm \ erg\ s^{-1}$ (see Table 2), which is about 1/3 of that given by
\citet{bott08}. The electrons distribution index used by
\citet{bott08} is $n=1.5$, which is almost the same as that we
obtained ($n\approx1.49$). However, in the framework of shock
acceleration, such a small electrons distribution index can only
result from the extreme condition \citep{Stecker07}, while in the
stochastic acceleration model, such a small one can be naturally
obtained assuming $t_{\rm acc}\sim t_{\rm esc}$. In order to
achieve the equipartition, a larger value ($\sim 23~\mu$G and $U^{\prime}_{\rm
B}\approx 0.1U^{\prime}_{\rm CMB}$, where $U^{\prime}_{\rm
CMB}=4.02\times10^{-13}(1+z)^4\delta^2_{\rm D}\ {\rm erg\
cm^{-3}}$) of magnetic field strength is required in our model
compared to that ($\sim 10~\mu$G) used in
\citet{bott08}, which causes the larger synchrotron emission from
extended jet than that derived in \citet{bott08}. For the inner
jet, we assumed a broken power-paw electrons distribution to
create the non-thermal emission. It can be found that the values
of $B^{\prime}$,$R^{\prime}_{\rm b}$, $\delta_{\rm D}$ and
$\gamma^{\prime}_{\rm max}$ used in our model are very close to
those used by \citet{bott08}. However, the value of
$\gamma^{\prime}_{\rm min}$ can be as low as one in our inner jet
model ($\gamma^{\prime}_{\rm min} \sim 10^5$ in \citet{bott08}).

\section{Discussion and conclusion}

Recently, some studies indicate that particles in the large-scale
jets of radio galaxies would be accelerated to high energies
through the stochastic acceleration (e.g.,
\cite{Hardcastle,O'Sullivan}). Actually,  the stochastic
acceleration has been applied to explain emissions from the inner
jets of blazars (e.g., \cite{wang,kata06b,Tramacere11}). It is
likely that several acceleration mechanisms may be acting in
parallel in the jets of AGNs (e.g., \cite{Rieger}). Because of the
detections of X-ray and GeV emissions from the extended jets of
blazars and radio galaxies (e.g., \cite{abdo,Massaro}), we think
it is worthwhile to investigate whether TeV emission could be
created in the extended jet of a blazar. In this paper, we assume
that electrons in the extended jet of 1ES 1101-232 can be
accelerated to high energies by stochastic acceleration, therefore
TeV emission may be produced via Compton up-scattering of CMB and
EBL photons by these electrons. The traditional emission from the
inner jet should also be taken into account. For simplicity, we
use the conventional SSC model to produce this emission from the
inner jet. It can be seen that our model can well reproduce the
SED of 1ES 1101-232, and its hard TeV emission is dominated by the
IC/CMB component from the extended jet. Since our model requires
the extent of the emission region in the extended jet to be kiloparsec,
the behavior of substantial TeV variability is forbidden. The emissions
from a compact region in the inner jet make dominant contribution
to the emission of 1ES 1101-232 at optical- X-ray and GeV band.
Therefore, the variabilities on time-scales of days at optical -
X-ray and GeV band are allowed. If it is determined that the
extended jet is another TeV emission region, this fact will
provide stronger constraints on the physics properties of emission
region in the inner jet. However, it should be kept in mind that
this model is only applicable for the TeV blazars that show no TeV
variability (e.g., 1ES 1101-232, 1ES 0229+200 and 1ES 0347-121).

We assumed three characteristic values of Doppler factor for the
TeV emission region in the extended jet. We can well reproduce the
TeV emission of 1ES 1101-232 using each one of them. Furthermore,
for a moderate or a large value of Doppler factor ($\delta_{\rm
D}\sim10, 15$), the physical parameters in the TeV emitting region
in the extended jet are consistent with the equipartition (or quasi-equipartition). In
contrast, for a small value of $\delta_{\rm D}\sim5$, the parameters
largely depart from the equipartition with $U^{\prime}_{\rm
e}/U^{\prime}_{\rm B}\gg1$. \citet{Piner10} suggested that the
bulk Lorentz factors in the TeV emission region and radio emission
region of TeV blazar would be different. Therefore, we conclude,
from  our results, that the TeV emitting region in the extended
jet of 1ES 1101-232 should be moderately or highly beamed.

\section*{Acknowledgments}
We thank the anonymous referee for his/her very constructive
comments. This work is partially supported by  a 973 Program
(2009CB824800) and Yunnan Province under a grant 2009 OC.


\bibliography{refernces}

\begin{thebibliography}{}

\bibitem[Abdo et al.(2010)]{abdo}Abdo, A., A., et al., 2010, Sci, 328, 725

\bibitem[Abdo et al.(2010)]{abdoagn}Abdo, A., A., et al., 2010, ApJ, 715, 429

\bibitem[Ackermann et al.(2011)]{Ackermann}Ackermann, M., et al., 2011, ApJ, 743, 171

\bibitem[Aharonian et al.(2006)]{Aharonian06}Aharonian, F., et al. 2006, Nature, 440, 1018

\bibitem[Aharonian et al.(2007a)]{Aharonian07a}Aharonian, F.,  et al., 2007a, A\&A, 475, L9

\bibitem[Aharonian et al.(2007b)]{Aharonian07b}Aharonian, F.,  et al., 2007b, A\&A, 473, L25

\bibitem[Aharonian et al.(2007c)]{Aharonian07c}Aharonian, F.,  et al., 2007c, A\&A, 470, 475

\bibitem[Aharonian et al.(2008)]{Aharonian08}Aharonian, F. A., Khangulyan, D., \& Costamante, L. 2008, MNRAS, 387, 1206

\bibitem[Atoyan \& Dermer(2004)]{Atoyan}Atoyan, A., \& Dermer, C. D. 2004, ApJ, 613, 151

\bibitem[B\"{o}ttcher et al.(1997)]{bott97}B\"{o}ttcher, M., Mause, H., \& Schlickeiser, R. 1997, A\&A, 324, 395

\bibitem[B\"{o}ttcher et al.(2008)]{bott08}B\"{o}ttcher, M., Dermer, C. D., \& Finke, J. D., 2008, ApJ, 679, L9

\bibitem[B\"{o}ttcher(2007)]{bott07} B\"{o}ttcher,  M. 2007, Ap\&SS, 309, 95

\bibitem[Celotti \& Fabian(1993)]{Celotti}Celotti, A., \& Fabian, A. C. 1993, MNRAS, 264, 228

\bibitem[Celotti \& Ghisellini(2008)]{Celotti08}Celotti, A., \& Ghisellini, G. 2008, MNRAS, 385, 283

\bibitem[Costamante(2011)]{Costamante2011}Costamante, L. 2011 Fermi Symposium, Roma.

\bibitem[Chang \& Copper(1970)]{cc}Chang, J. S., \& Copper, G. 1970, J. Comput. Phys., 6, 1

\bibitem[Dom\'{\i}nguez et al.(2011)]{Dom2011}Dom\'{\i}nguez, A. et al., 2011, MNRAS, 410, 2556

\bibitem[Dermer \& Schlickeiser(1993)]{dermer93}Dermer, C. D., \& Schlickeiser, R. 1993, ApJ, 416, 458

\bibitem[Dermer et al.(2009)]{dermer09}Dermer, C. D., Finke,  J. D., Krug,  H., \& B\"{o}ttcher,  M. 2009, ApJ, 692, 32

\bibitem[Essey et al.(2011)]{Essey}Essey, W., Kalashev, O., Kusenko, A., \& Beacom, J. F. 2011, ApJ, 731, 51

\bibitem[Fan et al.(2008)]{fan}Fan, Z. H., Liu, S. M., Wang, J. M., Fryer, C. L., \& Li, H. 2008, ApJ, 673, L139

\bibitem[Finke et al.(2008)]{finke08}Finke, J. D., Dermer, C. D., \& B\"{o}ttcher, M. 2008, ApJ, 686,
181

\bibitem[Finke et al.(2010)]{finke10}Finke, J. D., Razzaque,  S., \& Dermer, C. D.
2010, ApJ, 712, 238

\bibitem[Finke et al.(2012)]{finke12}Finke, J. D., Georganopoulos, M., \& Reyes, L. for the Fermi-LAT Collaboration, "Constraints on the Extragalactic Background Light from Gamma-Ray Observations", SnowTIGER 2012, Snowbird, Utah, March 2012

\bibitem[Franceschini et al.(2008)]{Franceschini}Franceschini, A., Rodighiero, G., \& Vaccari, M. 2008, A\&A, 487, 837

\bibitem[Georganopoulos et al.(2008)]{Georganopoulos}Georganopoulos, M. et al., 2008, ApJ, 686, L5

\bibitem[Ghisellini \& Tavecchio(2009)]{G09}Ghisellini, G. \& Tavecchio, F., 2009, MNRAS, 397, 985

\bibitem[Hardcastle et al.(2008)]{Hardcastle}Hardcastle, M. J., Cheung C. C., Feain I. J., Stawarz L., 2008, MNRAS, 393,
1041

\bibitem[Hardcastle \& Croston(2011)]{Hardcastle11}Hardcastle, M. J., \& Croston, J. H., 2011, MNRAS, 415, 133

\bibitem[Jorstad \& Marscher(2004)]{Jorstad}Jorstad, S. G., \& Marscher, A. P., 2004, ApJ, 614, 615

\bibitem[Jones(1968)]{Jones68}Jones, F. C. 1968, Phys. Rev., 167, 1159

\bibitem[Katarzy\'{n}ski et al.(2006a)]{kata06a}Katarzy\'{n}ski, K., Ghisellini, G., Tavecchio, F., Gracia, J., \& Maraschi, L. 2006a, MNRAS, 368, L52

\bibitem[Katarzy\'{n}ski et al.(2006b)]{kata06b}Katarzy\'{n}ski, K., Ghisellini, G., Mastichiadis, A., Tavecchio, F., \& Maraschi, L.
2006b, A\&A, 453, 47

\bibitem[Kirk et al.(1998)]{Kirk}Kirk, J. G., Rieger, F. M., \& Mastichiadis, A. 1998, A\&A, 333, 452

\bibitem[Laurent-Muehleisen et al.(1993)]{LM}Laurent-Muehleisen, S. A., Kollgaard, R. I., Moellenbrock, G. A., \& Feigelson, E. D. 1993, AJ, 106, 875

\bibitem[Lefa et al.(2011a)]{lefa11a}Lefa, E., Rieger, F. M., \& Aharonian, F. A. 2011a, ApJ, 740, 64

\bibitem[Lefa et al.(2011b)]{lefa11b}Lefa, E., Aharonian, F., \& Rieger, F. M.  2011b, ApJ, 743, L19

\bibitem[Mannheim(1993)]{Mannheim}Mannheim, K., 1993, A\&A, 269, 67

\bibitem[Maraschi et al.(1992)]{Maraschi}Maraschi, L., Ghisellini, G., \& Celotti, A. 1992, ApJL, 397, L5

\bibitem[Massaro et al.(2011)]{Massaro}Massaro, F., Harris, D. E., \& Cheung, C. C., 2011, ApJS, 197, 24

\bibitem[Moderski et al.(2005)]{Moderski05}Moderski, R., Sikora, M., Coppi, P. S., \& Aharonian, F. 2005, MNRAS, 363,
954

\bibitem[M\"{u}cke et al.(2003)]{Mucke}M\"{u}cke, A., Protheroe, R. J., Engel, R. et al. 2003, APh, 18, 593

\bibitem[Neronov \& Vovk (2010)]{Neronov}Neronov, A.,\& Vovk, I., 2010, Sci, 328, 73

\bibitem[O'Sullivan et al.(2009)]{O'Sullivan}O'Sullivan, S., Reville, B., \& Taylor, A. M. 2009 MNRAS, 400, 248

\bibitem[Park \& Petrosian(1996)]{Park}Park, B. T., \& Petrosian, V. 1996, ApJS, 109, 255

\bibitem[Piner et al.(2008)]{Piner08}Piner, B. G., Pant, N., Edwards, P. G., 2008, ApJ, 678, 64

\bibitem[Piner et al.(2010)]{Piner10}Piner, B. G., Pant, N., Edwards, P. G., 2010, ApJ, 723, 1150

\bibitem[Rees(1967)]{rees}Rees, M. J. 1967, MNRAS, 137, 429

\bibitem[Remillard et al.(1989)]{Remillard}Remillard, R. A., Tuhoy, I. R., Brissenden, R. J. V., et al. 1989, ApJ, 345, 140

\bibitem[Rieger et al.(2007)]{Rieger}Rieger, F., Bosch-Ramon, V., Duffy, P., 2007, Ap\&SS, 309, 119

\bibitem[Romero et al.(1999)]{Romero}Romero, G. E., Cellone, S. A., \& Combi, J. A. 1999, A\&AS, 135, 477

\bibitem[Sambruna et al.(2008)]{Sambruna}Sambruna, R. M., Donato, D., Cheung, C. C., Tavecchio, F., Maraschi, L. 2008, ApJ, 684, 862

\bibitem[Stecker et al.(1992)]{Stecker92}Stecker, F. W., de Jager, O. C., \& Salamon, M. H. 1992, ApJ, 390, L49

\bibitem[Stecker et al.(2007)]{Stecker07}Stecker, F. W., Baring, M. G., \& Summerlin, E. J. 2007, ApJ, 667, L29

\bibitem[Sikora et al.(1994)]{Sikora94}Sikora, M., Begelman, M. C., \& Rees, M. J. 1994, ApJ, 421, 153

\bibitem[Tavecchio et al.(2000)]{tavecc00}Tavecchio, F., Maraschi, L., Sambruna, R. M., \& Urry, C. M., 2000, ApJ, 544,
L23

\bibitem[Tavecchio et al.(2007)]{tavecc07}Tavecchio, F., Maraschi, L., Wolter, A., Cheung, C. C., Sambruna, R. M., Urry, C. M. 2007, ApJ, 662, 900

\bibitem[Tavecchio et al.(2009)]{tavecc09}
Tavecchio, F., Ghisellini, G., Ghirlanda,  G., Costamante, L., \&
Franceschini, A. 2009, MNRAS, 399, 59

\bibitem[Tramacere et al.(2011)]{Tramacere11}Tramacere, A., Massaro, E.,\& Taylor, A. M. 2011, ApJ, 739, 66

\bibitem[Urry \& Padovani(1995)]{Urry}Urry, C. M., \&  Padovani, P., 1995, PASP, 107, 803

\bibitem[Wang(2002)]{wang}Wang, J. C., 2002, ChJAA, 2, 1

\bibitem[Weidinger et al.(2010)]{Weidinger10}Weidinger, M., R\"{u}ger, M., \& Spanier, F., 2010, Astrophys. Space Sci. Trans., 6, 1

\bibitem[Weidinger(2011)]{Weidinger11}Weidinger, M., PhD thesis, 2011, University of Wuerzburg

\bibitem[Wolter et al.(2000)]{Wolter}Wolter, A., Tavecchio, F., Caccianiga, A., Ghisellini, G., \& Tagliaferri, G., 2000, A\&A, 357, 429

\bibitem[Zacharopoulou et al.(2011)]{Zach}Zacharopoulou, O., Khangulyan, D., Aharonian, F., \& Costamante, L.
2011, ApJ, 738, 157

\end{thebibliography}

\end{document}